\documentclass[aps,superscriptaddress,amsfonts,amssymb,amsmath,showpacs,nofootinbib, twocolumn]{revtex4-1}
\usepackage[dvips]{graphicx}

\begin{document}

\title{The effect of magnetic fields on the r-modes of slowly rotating\\ relativistic neutron stars}

\author{Cecilia Chirenti} 
\email{e-mail: cecilia.chirenti@ufabc.edu.br} 
\affiliation{Centro de Matem\'atica, Computa\c c\~ao e Cogni\c c\~ao, UFABC, 09210-170 Santo Andr\'e, SP, Brazil}

\author{Jozef Sk\'akala}
\email{e-mail: jozef.skakala@ufabc.edu.br}
\affiliation{Centro de Matem\'atica, Computa\c c\~ao e Cogni\c c\~ao, UFABC, 09210-170 Santo Andr\'e, SP, Brazil}

\begin{abstract}
We study here the r-modes in the Cowling approximation of a slowly rotating and magnetized neutron star with a poloidal magnetic field, where we neglect any deformations of the spherical symmetry of the star. We were able to quantify the influence of the magnetic field in both the oscillation frequency $\sigma_r$ of the r-modes and the growth time $t_{gw}$ of the gravitational radiation emission. We conclude that magnetic fields of the order $10^{15}$ G at the center of the star are necessary to produce any changes. Our results for $\sigma_r$ show a decrease of up to $\sim$ 5\% in the frequency with increasing magnetic field, with a $B^2$ dependence for rotation rates $\Omega/\Omega_K \gtrsim 0.07$ and $B^4$ for $\Omega/\Omega_K \lesssim 0.07$. (These results should be trusted only within slow rotation approximation and we kept $\Omega/\Omega_{K}< 0.3$.)  For $t_{gw}$, we find that it is approximately 30\% smaller than previous Newtonian results for non-magnetized stars, which would mean a faster growth of the emission of gravitational radiation. The effect of the magnetic field in $t_{gw}$ causes a non-monotonic effect, that first slightly increases $t_{gw}$ and then decreases it further by another $\sim$ 5\%. (The value of magnetic field for which $t_{gw}$ starts to decrease depends on the rotational frequency, but it is generally around $10^{15}$G.)  Future work should be dedicated to the study of the effect of viscosity in the presence of magnetic fields, in order to establish the magnetic correction to the instability window. 
\end{abstract}

\maketitle

\section{Introduction}
The r-mode instability was first discovered in \cite{Andersson-instability, Friedman-instability} and it was predicted that the instability could become a significant source of gravitational radiation. This happens because the r-modes are generically unstable to the CFS gravitational-radiation-driven instability \cite{Chandrasekhar,Friedman-Schutz1,Friedman-Schutz2}. The r-mode instability follows immeadiately from the fact that r-modes that are prograde with respect to a distant observer are retrograde in the comoving frame for all values of the angular velocity (the canonical energy of the modes is negative). For some nice reviews see \cite{Kokkotas3, Stergioulas}. However, different mechanisms to damp the instability have to be considered: one of them is viscosity \cite{Lindblom}, another one could be sufficiently strong magnetic fields \cite{Rezzolla1, Rezzolla2, Rezzolla3}. The instability windows for non-magnetized Newtoninan stars were initially calculated in \cite{Lindblom2, Andersson}. More recently, the effect of magnetic fields on r-modes was discussed in \cite{Rezania, Abbassi} for a spherical shell and in \cite{Lander} for a neutron star with purely toroidal field, in all cases in the Newtonian context. 

This paper is a first part of a project that is supposed to contribute to the understanding of the instability window for slowly rotating relativistic neutron stars with magnetic fields. We are here interested in and focused on the modification of the r-modes in the presence of magnetic fields and its effects on the gravitational wave emission. The effect of magnetic fields on the r-mode frequencies could be interesting for astrophysical objects such as magnetars, that have magnetic fields of the order of magnitude $10^{15}$ G and are very slow rotators with rotational periods of a few seconds. We consider here stars of comparable magnetic fields with rotational periods of a few milliseconds (due to a numerical difficulty: longer periods would need longer time evolutions). The issue of viscous damping of the mode, which determines the instability window of the r-mode, is further complicated by the presence of the magnetic field. Therefore we leave it for future work.

 We treat the problem within the realm of perturbation theory, first by deriving general perturbation equations and then by solving them numerically with a 2D Lax-Wendroff method. The same numerical methods were used and tested in our previous paper \cite{Chirenti}. The advantages of the 2D dynamical evolution in this case is that it avoids both the r-mode continuous spectrum problem \cite{Beyer} and the need for truncating the solution at some $\ell$ (as done for instance in \cite{Kokkotas1, Kokkotas2} for torsional modes of a relativistic star with a dipole magnetic field). We first calculate the r-mode frequencies for different values of the rotation parameter and magnetic fields. Then we calculate the instability growth rate due to the emission of gravitational waves as a function of the magnetic field. (This gives both general relativistic and magnetic field corrections to the results of \cite{Lindblom2, Andersson}.)

The paper is organized as follows: in the second section we describe our background model and in the third section we present the full perturbation equations of our model. This section is followed by the fourth section, where we compute the r-mode frequencies for different rotation rates of the star, as a function of the magnetic field. In the fifth section we compute the growth time due to the r-mode gravitational wave emission as a function of magnetic field. In the sixth section we present the concluding remarks. (Everywhere in the paper, unless explicitly mentioned, we use the units $c=G=M_{\bigodot}=1$.)

\section{The background model}

We work with a slowly and uniformly rotating magnetized star with a polytropic equation of state. Our model neutron star has $M = 1.4 M_{\bigodot}$, $R = 14.08$ km, and the pressure $p$ is given by the polytropic equation of state $p=K\rho^{\Gamma}$, taken with the parameters $K=100$, $\Gamma=2$ and $\rho$ is the rest-mass density of the star. The Keplerian frequency $\Omega_{K}$ (mass shedding limit) that we use in our paper to normalize the rotation of the star is ~$\Omega_{K}=\sqrt{M/R^{3}}=1.3$ \textrm{kHz}. 

The effect of the rotation is taken up to the linear order in the rotation parameter $\Omega$, which means one considers the effect of the rotation on the spacetime metric (frame dragging function), but neglects the effect of the rotation on the stellar structure. (The deformations of the stellar fluid due to rotation are of the order $\Omega^{2}$.)

We consider a dipole magnetic field and for the realistic neutron stars with magnetic fields (up to the order of $10^{15}G$ for magnetars), one can neglect the effect of the magnetic field on both the stellar structure and the background metric (for a more detailed argumentation, see \cite{Kokkotas1}). 

 This means the model follows from a line element:
\[ ds^{2}=-e^{\nu}dt^{2}+e^{\lambda}dr^{2}+r^{2}d\theta^{2}+r^{2}\sin^{2}(\theta)\left[d\phi-\omega dt\right]^{2},\]
where $\nu$, $\lambda$ and $\omega$  are functions of $r$, and a stress energy tensor given by:
\[ T^{\mu\nu}=(p+\epsilon+H^{2})u^{\mu}u^{\nu}+\left(p+\frac{H^{2}}{2}\right)g^{\mu\nu}-H^{\mu}H^{\nu}. \]
Here $\epsilon$ is the total energy density, the 4-velocity $u^{\mu}$ is given as 
\[ u^{(t,r,\theta,\phi)}=\left(e^{-\nu/2},~0,~0,~\Omega e^{-\nu/2}\right),\]
and $H^{\mu}$ is related to the magnetic field $B^{\mu}$, defined as usual in terms of the electromagnetic tensor $F^{\mu\nu}$ as
\[H_{\mu}=\frac{B_{\mu}}{4\pi}=-\frac{1}{8\pi}\epsilon_{\mu\nu\alpha\beta}u^{\nu}F^{\alpha\beta}.\] 
Our background model is thus obtained by the TOV equations supplemented by a polytropic equation of state, Hartle's equation for the frame dragging function $\omega$ \cite{Hartle}:
\begin{equation}
\frac{e^{\frac{\nu+\lambda}{2}}}{r^4}\left( r^4 e^{-\frac{\nu+\lambda}{2}}\omega_{,r}\right)_{,r} + \frac{2}{r}(\nu_{,r} + \lambda_{,r})(\Omega - \omega) = 0\,,
\label{eq:Hartle}
\end{equation}
and the magnetic field is given by the Maxwell equations. Therefore our dipole equilibrium magnetic field (for our choice of equilibrium magnetic field, the induction equation is  trivially fulfilled) obeys the Maxwell equations
\[F^{\mu\nu}_{~;\nu}=4\pi J^{\mu},\]
solved with a 4-current $J_{\mu}$ with the only non-zero component
\begin{equation}
\label{eq:J_mu}
4\pi J_{\phi}(r,\theta) = -\alpha(r)\sin^2(\theta)\,,
\end{equation}
(this choice is equivalent, for instance, to choosing the terms with parity $(-1)^{\ell+1}$ in the parity decomposition given in \cite{Ruffini} and keeping only $\ell = 1$, $m = 0$) with the radial profile 
\[ \alpha (r) = \alpha_{0} r^{2}\epsilon^{2}(r)~,   \]
that describes a ring current inside the star. Moreover, this current profile allows us to consider the magnetic field as force-free at the surface (the Lorentz force goes to zero at the surface of the star as $\epsilon^2$).

Choosing the vector potential as $A_{\phi} = -a(r)\sin^2(\theta)$ (same parity and symmetry choices as we did for $J^{\mu}$ above), the Maxwell equations give:

\begin{equation}
\label{eq:a_gen}
e^{-\lambda}a_{,rr} + \left[ 4\pi(p-\epsilon) r +
  \frac{1-e^{-\lambda}}{r}\right]a_{,r} - 
\frac{2}{r^2}a + \alpha = 0.
\end{equation}
Here the components of our dipole magnetic field are obtained from $a(r)$ as:
\[ H^{r}(r,\theta)=\frac{a(r)}{2\pi r^{2}}\cos \theta~,\]
\[ H^{\theta}(r,\theta)=\frac{a_{,r}e^{-\lambda}}{4\pi r^{2}}\sin\theta~.\]
It is known that the equation \eqref{eq:a_gen} has outside the star an exact analytic solution \cite{Wasserman}: 
\begin{equation} 
a(r)= C r^{2}\left[\ln\left(\frac{r}{r-2M}\right)-\frac{2M (r+M)}{r^{2}}\right].
\label{eq:a_exact}
\end{equation}
The full solution inside and outside the star for the dipole magnetic field was computed by numerically solving eq. (\ref{eq:a_gen}) inside the star, matching the regular series expansion of the solution near the center with the numerical solution up to the surface, where we require continuity of $a(r)$ and its first derivative with the exterior analytic solution (\ref{eq:a_exact}). 

\begin{figure}[!htb]
\begin{center}
\includegraphics[angle=270,width=1\linewidth]{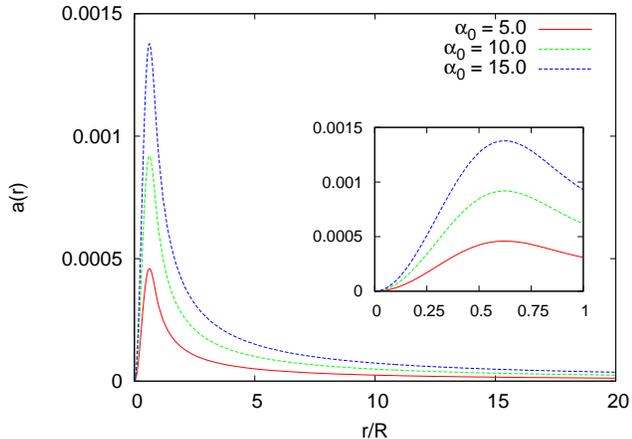}
\end{center}
\caption{The function $a(r)$ for different currents as a function of the radial coordinate divided by the radius of the star.}
\label{fig:teste_a}
\end{figure}

\begin{figure}[!htb]
\begin{center}
\includegraphics[angle=270,width=1\linewidth]{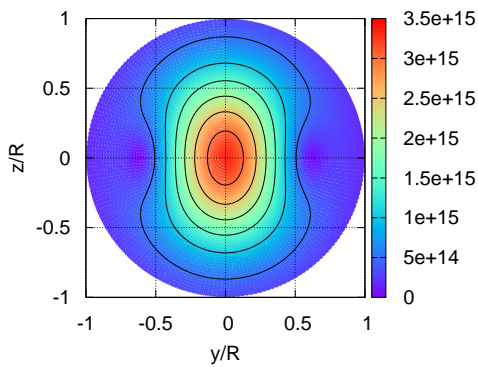}
\end{center}
\caption{The absolute value of the dipole magnetic field (in Gauss) corresponding to $\alpha_{0}=10$. The magnetic field is shown in the plane given by the rotational axis and a perpendicular direction to the rotational axis.}
\label{fig:teste_B2}
\end{figure}

\begin{figure}[!htb]
\begin{center}
\includegraphics[angle=270,width=1\linewidth]{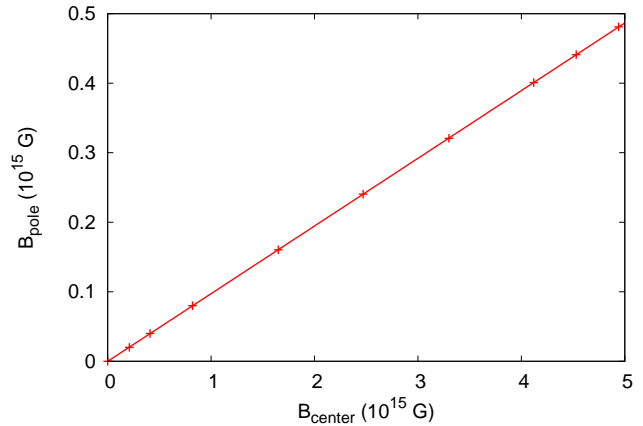}
\end{center}
\caption{The value of the magnetic field at the magnetic pole $B_{\textrm{pole}}$ as a funtion of the magnetic field at the center of the star $B_{\textrm{center}}$ as calculated with our model. We obtained $B_{\textrm{pole}} = 0.0974 B_{\textrm{center}}$.}
\label{fig:teste_Bpole}
\end{figure}

Some representative plots of the behavior of the radial function $a(r)$ can be seen in figure \ref{fig:teste_a}, for increasing values of the current parameter $\alpha_0$. One typical solution for the amplitude of the magnetic field inside the star is given in figure \ref{fig:teste_B2} (we note here that the solid lines are contour lines, and not magnetic field lines). Throughout the paper we refer to the value of the magnetic field at the center of the star. But since the magnetic field at the pole of a star can be determined observationally, while the value of the magnetic field at the center of star must be calculated with some model, we present in figure \ref{fig:teste_Bpole} the relation between $B_{\rm{pole}}$ and $B_{\rm{center}}$ given by our model.

\begin{figure}[!htb]
\begin{center}
\includegraphics[angle=270,width=1\linewidth]{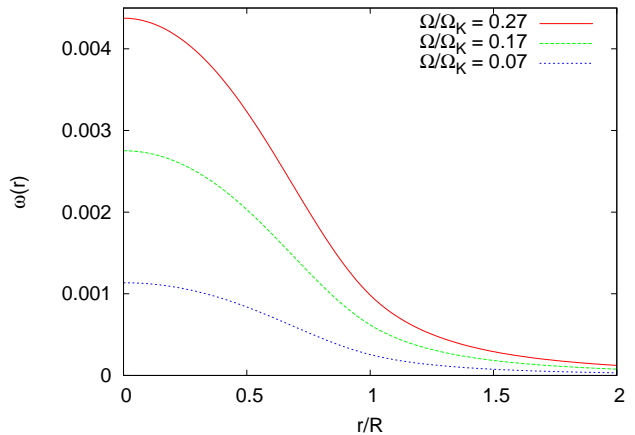}
\end{center}
\caption{The frame dragging function $\omega$ as a function of the radial coordinate normalized by the radius of the star. (The frame dragging function up to the linear order in $\Omega$ is for a uniformly rotating star only a function of $r$ and outside the star behaves as $2J/r^{3}$ \cite{Hartle}.)}
\label{fig:teste_omee}
\end{figure}

In figure \ref{fig:teste_omee} we present some representative plots of the numerical solutions obtained for the frame dragging function $\omega$, from Hartle's equation (\ref{eq:Hartle}).

\section{Perturbation equations}
We are working in the Cowling approximation and considering only barotropic perturbations, therefore the fundamental set of perturbation variables is given by ~$\delta p, \delta u^{\nu}, \delta H^{\nu}$. (For an analysis of the accuracy of the Cowling approximation, but only for f and p modes, see \cite{Kojima}. For r-modes, the Cowling approximation gives more accurate frequencies, as these modes do not involve large density variations \cite{Friedman}.)
The perturbation equations are obtained by perturbing the Euler equations ($i=r,\theta,\phi$):
\begin{equation}
\delta \left( \{\delta^{i}_{\nu}+u^{i}u_{\nu}\}T^{\nu\beta}_{~;\beta}\right)=0~
\end{equation}
and the energy conservation equation
\begin{equation}
\delta\left(u_{\nu}T^{\nu\beta}_{~;\beta}\right)=0~
\end{equation}
together with the perturbed induction equations
\begin{equation}
\delta\left\{(u^{\mu}H^{\nu}-H^{\mu}u^{\nu})_{;\nu}\right\}=0.
\end{equation}
Furthermore one uses as constraints the perturbed ideal MHD equation:
\begin{equation}
\delta (H^{\mu}_{;\mu})=\delta \left(H_{\mu}u^{\mu}_{;\nu}u^{\nu}\right)~,
\end{equation} 
and the 4-velocity normalization condition
\[\delta (u^{\nu}u_{\nu})=0~,\]
together with the fact that the magnetic field remains perpendicular to the 4-velocity
\[\delta (u_{\nu}H^{\nu})=0~.\]
The last two equations can be subsequently used to reduce the variables to the 7 independent variables ~$\delta p, \delta H^{i}, \delta u^{i}$,
with ~$i=r,\theta,\phi$. 

\subsection{The original form of the perturbation equations}

There are 3 independent components of the perturbed induction equation which turn into:
\begin{widetext}
\begin{itemize}
\item the $r$ component:
\begin{eqnarray}
e^{-\nu/2}(\delta H^{r}_{,t}+\Omega\delta H^{r}_{,\phi})+(H^{r}_{,\theta}+H^{r}\cot(\theta))\delta u^{\theta}+H^{r}(\delta u^{t}_{,t}+\delta u^{\phi}_{,\phi}+\delta u^{\theta}_{,\theta})-~~~~~~~~~~~~~~~~~~~~~~~~~~~~~~~~~~~~\nonumber\\
\left(H^{\theta}_{,\theta}+\cot(\theta)H^{\theta}\right)\delta u^{r}-H^{\theta}\delta u^{r}_{,\theta}=0.~~~~~~~~~~~~~~~~~~~~~~~~~~~~~~~~~~~~ 
\end{eqnarray}

\item the $\theta$ component:

\begin{eqnarray}
e^{-\nu/2}(\delta H^{\theta}_{,t}+\Omega \delta H^{\theta}_{,\phi})+H^{\theta}(\delta u^{t}_{,t}+\delta u^{r}_{,r}+\delta u^{\phi}_{,\phi})+~~~~~~~~~~~~~~~~~~~~~~~~~~~~~~~~~~~~~~~~~~~~~~~~~~~~~~~~~~~~~~~~\nonumber\\
+\left\{H^{\theta}_{,r}+H^{\theta}\left(\frac{\nu_{,r}+\lambda_{,r}}{2}+\frac{2}{r}\right)\right\}\delta u^{r}-\left(H^{r}_{,r}+\left\{\frac{\nu_{,r}+\lambda_{,r}}{2}+\frac{2}{r}\right\}H^{r}\right)\delta u^{\theta}-H^{r}\delta u^{\theta}_{,r}=0.~~~~~~~~~~~~~~~~~~~~~~~~~~~ 
\end{eqnarray}

\item the $\phi$ component:

\begin{eqnarray}
e^{-\nu/2}\delta H^{\phi}_{,t}-\frac{\nu_{,r}}{2}H^{r}\delta u^{\phi}-~~~~~~~~~~~~~~~~~~~~~~~~~~~~~~~~~~~~~~~~~~~~~~~~~~~~~~~~~~~~~~~~~~~~~~~~~~~~~~~~~~~~\nonumber\\
H^{r}\delta u^{\phi}_{,r}-H^{\theta}\delta u^{\phi}_{,\theta}-\Omega e^{-\nu/2}(\delta H^{t}_{,t}+\delta H^{r}_{,r}+\delta H^{\theta}_{,\theta}+\left(\frac{\lambda_{,r}}{2}+\frac{2}{r}\right)\delta H^{r}+\cot(\theta)\delta H^{\theta})=0.~~~~~~~~~~~~~~~~~~~~~~~~~~~~~~~~~ 
\label{induc_phi}
\end{eqnarray}
\end{itemize}

The perturbed energy conservation equation is independent of the magnetic fields and is given as:
\begin{eqnarray}\label{4}
\delta p_{,t}+\Omega\cdot\delta p_{,\phi}+e^{\nu/2}\cdot\Gamma p\left[e^{-\nu}r^{2}\sin^{2}(\theta)\cdot(\Omega-\omega)\cdot\delta u^{\phi}_{,t}+\delta u^{r}_{,r}+\delta u^{\theta}_{,\theta}+\delta u^{\phi}_{,\phi}\right]=~~~~~~~~~~~~~\nonumber\\
=-e^{\nu/2}\cdot\left\{\Gamma p\left[\frac{\nu_{,r}+\lambda_{,r}}{2}+\frac{2}{r}\right]-(p+\epsilon)\frac{\nu_{,r}}{2}\right\}\cdot\delta u^{r}-e^{\nu/2}\cdot\Gamma p\cdot\cot(\theta)\cdot\delta u^{\theta}~.~~~~~~~~
\end{eqnarray}

The independent components of the perturbed Euler equation:
\begin{itemize}
\item the $r$ component
\begin{eqnarray}
(\epsilon+p+H^{r2}e^{\lambda}+H^{\theta 2}r^{2})\{e^{-\nu/2}\Omega\delta u^{r}_{,\phi}+e^{-\nu/2-\lambda}r^{2}\sin^{2}(\theta) [\omega_{,r}+  \left(  \nu_{,r}-\frac{2}{r}  \right)  (\Omega-\omega)]\delta u^{\phi}\}+~~~~~~~~~~~~~~~~~~~~~~~~~~~~~\nonumber\\
  \left[  (\epsilon+p+H^{r2}e^{\lambda}+H^{\theta 2}r^{2})e^{-\nu/2}-e^{\lambda-\nu/2}H^{r 2}  \right]  \delta u^{r}_{,t}\frac{e^{-\lambda}\nu_{,r}}{2}(\delta\epsilon+\delta p)=~~~~~~~~~~~~~~~~~~~~~\nonumber\\
=-e^{-\lambda}\delta p_{,r}+H^{r}(\delta H^{r}_{,r}+e^{-\nu/2}r^{2}H^{\theta}\delta u^{\theta}_{,t}+e^{-\nu}r^{2}\sin^{2}(\theta)(\Omega-\omega)\delta H^{\phi}_{,t}+\delta H^{\phi}_{,\phi}+\delta H^{\theta}_{,\theta})+~~~~~~~~~~~~~~~~~~~\\
+\delta H^{\theta}  \left(  -e^{-\lambda}r^{2}H^{\theta}  \left[  \frac{\nu_{,r}}{2}+\frac{2}{r}  \right]  +\cot(\theta)H^{r}  \right)  +\delta H^{r}_{,\theta}H^{\theta}-e^{-\lambda}r^{2}H^{\theta}\delta H^{\theta}_{,r}+~~~~~~~~~~~~~~~~~~~~\nonumber\\
+\delta H^{r} H^{r}  \left\{  \frac{\lambda_{,r}}{2}+\frac{2}{r}  \right\}~~~~~~~~~~~~~~~~~~~~~~~~~~~\nonumber
\end{eqnarray}

\item the $\theta$ component
\begin{eqnarray}
(\epsilon+p+H^{r2}e^{\lambda}+H^{\theta 2}r^{2})\{e^{-\nu/2}\Omega\delta u^{\theta}_{,\phi}-2e^{-\nu/2}\sin(\theta)\cos(\theta)\cdot(\Omega-\omega)\cdot\delta u^{\phi}\}+~~~~~~~~~~~~~~~~~~~~~~~~~~~~~~~~~~~~~~~~~\nonumber\\
+\left[(\epsilon+p+H^{r2}e^{\lambda}+H^{\theta 2}r^{2})e^{-\nu/2}-e^{-\nu/2}r^{2}H^{\theta 2}\right]\delta u^{\theta}_{,t}=~~~~~~~~~~~~~~~~~~~~~~~~~~~~~~~~~~~~~~~~~\nonumber\\
=-\frac{1}{r^{2}}\delta p_{,\theta}+\delta H^{\theta}\left[H^{\theta}\cot(\theta)+H^{r}\left(\frac{\nu_{,r}}{2}+\frac{2}{r}\right)\right]+~~~~~~~~~~~~~~~~~~~~~~~~~~~~\\
+\delta H^{r}H^{\theta}\left(\frac{\lambda_{,r}}{2}+\frac{2}{r}\right)+H^{r}\delta H^{\theta}_{,r}+~~~~~~~~~~~~~~~~~~~~~~~~~~~~\nonumber\\
+H^{\theta}(e^{\lambda-\nu/2}H^{r}\delta u^{r}_{,t}+e^{-\nu}r^{2}\sin^{2}(\theta)(\Omega-\omega)\delta H^{\phi}_{,t}+\delta H^{r}_{,r}+\delta H^{\theta}_{,\theta}+\delta H^{\phi}_{,\phi})-\frac{e^{\lambda}}{r^{2}}H^{r}\delta H^{r}_{,\theta}.~~~~~~~~~~~~~~~~~~~~~\nonumber
\end{eqnarray}

\item the $\phi$ component

\begin{eqnarray}
(\epsilon+p+H^{r2}e^{\lambda}+H^{\theta 2}r^{2})\{e^{-\nu/2}\delta u^{\phi}_{,t}+e^{-\nu/2}\Omega\delta u^{\phi}_{,\phi}\}+~~~~~~~~~~~~~~~~~~~~~~~~~~~~~~~~~~~~~~~~~~~~~~~~\nonumber\\
+(\epsilon+p+H^{r2}e^{\lambda}+H^{\theta 2}r^{2})e^{-\nu/2}\left[(\Omega-\omega)_{,r}+\left(\frac{2}{r}-\nu_{,r}\right)(\Omega-\omega)\right]\delta u^{r}+~~~~~~~~~~~~~~~~\nonumber\
\nonumber\\
+2(\epsilon+p+H^{r2}e^{\lambda}+H^{\theta 2}r^{2})e^{-\nu/2}\cot(\theta)
(\Omega-\omega)\delta u^{\theta}=~~~~~~~~~~~~~~~~~\nonumber\\
=-\left[\frac{1}{r^{2}\sin^{2}(\theta)}\delta p_{,\phi}+e^{-\nu}(\Omega-\omega)\delta p_{,t}\right]+~~~~~~~~~~~~~~~~~~~~~~~~~~~~~~~~~~~~~~~~~~~~~~~~~~~~~\\
+H^{r}\delta H^{\phi}_{,r}+H^{\theta}\delta H^{\phi}_{,\theta}-\delta H^{t}\left\{\left(\frac{2\omega}{r}+\omega_{,r}+\frac{\Omega\nu_{,r}}{2}+\nu_{,r}(\Omega-\omega)\right)H^{r}+2\omega\cot(\theta)H^{\theta}\right\}+~~~~~~~~~~~~~~~~~~~~~~~~\nonumber\\
+\delta H^{\phi}\left(\left[\frac{\nu_{,r}}{2}+\frac{2}{r}\right]H^{r}+2\cot(\theta)H^{\theta}\right)-\Omega(H^{r}\delta H^{t}_{,r}+H^{\theta}\delta H^{t}_{,\theta})-
~~~~~~~~~~~~~~~~~~~~~~~\nonumber\\
\frac{e^{\lambda}}{r^{2}\sin^{2}(\theta)}H^{r}\delta H^{r}_{,\phi}-\frac{1}{\sin^{2}(\theta)}H^{\theta}\delta H^{\theta}_{,\phi}-e^{-\nu}(\Omega-\omega)(e^{\lambda}H^{r}\delta H^{r}_{,t}+r^{2}H^{\theta}\delta H^{\theta}_{,t}).~~~~~~~~~~~~\nonumber
\end{eqnarray}
\end{itemize}

The supplementary three constraints are the following:
\begin{itemize}
\item the perturbed ideal MHD equation
\begin{eqnarray}
e^{-\nu}r^{2}\sin^{2}(\theta)(\Omega-\omega)\delta H^{\phi}_{,t}+\delta H^{r}_{,r}+\delta H^{\theta}_{,\theta}+\delta H^{\phi}_{,\phi}+\left(\frac{\lambda_{,r}}{2}+\frac{2}{r}\right)\delta H^{r}+\cot(\theta)\delta H^{\theta}=H^{r}e^{\lambda-\nu/2}\Omega\delta u^{r}_{,\phi}+~~~~~~~~~~
\nonumber\\
+\delta u^{\phi}r^{2}\sin^{2}(\theta) e^{-\nu/2}\left(H^{r}\left[(\Omega-\omega)\left(\nu_{,r}-\frac{2}{r}\right)+\omega_{,r}\right]-2H^{\theta}\cot(\theta)(\Omega-\omega)\right)+H^{\theta}r^{2}e^{-\nu/2}\Omega\delta u^{\theta}_{,\phi}.~~~~~~~~~~~~~~~
\label{idealMHD}
\end{eqnarray}
\item perturbed perpendicularity condition of magnetic field and 4-velocity
\begin{equation}
e^{\nu/2}\delta H^{t}=e^{\lambda}H^{r}\delta u^{r}+r^{2}H^{\theta}\delta u^{\theta}+e^{-\nu/2}r^{2}\sin^{2}(\theta)(\Omega-\omega)\delta H^{\phi}~,
\end{equation}
\item and the perturbed 4-velocity normalization condition:
\begin{equation}
\delta u^{t}=e^{-\nu}r^{2}\sin^{2}(\theta)(\Omega-\omega)\delta u^{\phi}~.
\end{equation}
\end{itemize}

Let us mention that the upper equations reduce for the special case of~ $\Omega=\delta p=\delta u^{r}=\delta u^{\theta}=0$ to the equations shown in \cite{Kokkotas1}.

\end{widetext}

\subsection{The form of perturbation equations suitable for the numerical integration}

For the numerical integration with the 2D Lax-Wendroff scheme we need to obtain the dynamical equations in a form containing only one time derivative in each equation. This is most convenient to achieve by proper linear combinations of the original equations and ommiting the $\sim\Omega^{2}$ terms. The 7 independent variables ~$\delta H^{i}, \delta u^{i}, \delta p$~, ($i=r,\theta,\phi$),~ are further transformed into ``momentum-like'' variables
\[ \delta \tilde H^{i}=(\epsilon+p)\delta H^{i}, ~~~~ \delta \tilde u^{i}=(\epsilon+p)\delta u^{i}~.\]
(For the introduction of these variables in the Newtonian context see \cite{Jones}.) This transformation is done for the purpose of obtaining a simple boundary condition at the stellar surface as:
\[\delta \tilde u^{i}=\delta \tilde H^{i}=\delta p=0.\]

Furthermore we apply regularity conditions at the center of the star and at the rotational axis, together with the correct symmetry conditions at the equatorial plane. (For the details about these conditions see our previous work \cite{Chirenti}.) 

Another constraint that has to be fulfilled is the time independent MHD equation, that is checked to be satisfied in each step of the calculation (up to certain determined numerical error). The time independent MHD constraint is obtained from the perturbed ideal MHD equation \eqref{idealMHD} by subtracting the appropriate multiple of the $\phi$ component of the perturbed induction equation \eqref{induc_phi}. The constraint reads: 
\begin{widetext}
\begin{eqnarray}
e^{-\nu/2}r^{2}\sin^{2}(\theta)(\Omega-\omega)\left(H^{r}\delta u^{\phi}_{,r}+H^{\theta}\delta u^{\phi}_{,\theta}\right)+~~~~~~~~~~~~~~~~~~~~~~~~~~~~~~~~~~~~~~~~~~~~~~~~~~~~~~~~~~~\nonumber\\
+\delta H^{r}_{,r}+\delta H^{\theta}_{,\theta}+\delta H^{\phi}_{,\phi}+\left(\frac{\lambda_{,r}}{2}+\frac{2}{r}\right)\delta H^{r}+\cot(\theta)\delta H^{\theta}=H^{r}e^{\lambda-\nu/2}\Omega\delta u^{r}_{,\phi}+~~~~~~~~~~~~~~~~~~~~~~~~~~~~~~~~~~~~~~~~~~~~~~~~~~~~~~~~~~~~
\nonumber\\
+\delta u^{\phi}r^{2}\sin^{2}(\theta) e^{-\nu/2}\left(H^{r}\left[(\Omega-\omega)\left(\frac{\nu_{,r}}{2}-\frac{2}{r}\right)+\omega_{,r}\right]+2H^{\theta}\cot(\theta)(\omega-\Omega)\right)+H^{\theta}r^{2}e^{-\nu/2}\Omega\delta u^{\theta}_{,\phi}.~~~~~~~~~~~~~~~~~~~~~~~~~~~~~~~~~~~~~~~~~~~~~~~~~~~~~~\nonumber
\end{eqnarray}
\end{widetext}

Our integration domain occupies only the first quadrant, since we take advantage of the symmetries at the equatorial plane. Our numerical grid typically has $50 \times 50$ points in $r \times \theta$, where $r$ varies in $[0,R]$ and $\theta$ in $[0,\pi/2]$. We take usually 10.000-50.000 time steps in the evolution of the equations, depending on the rotating rate and, consequently, the frequency of the r-mode. In each time evolution we observe at least several periods of oscillation of the perturbations. We point out here that our time evolutions were so far stable, and we did not see signs of the hydromagnetic instability observed in \cite{LanJones} for axial-led perturbations in Newtonian gravity. The investigation of this issue in our relativistic treatment is left for a future work.

We limited the maximum rotation rate considered here by $\Omega = 0.27\Omega_K$, motivated by the results of \cite{Passamonti0}, where they see corrections of the order $\Omega^3$ in the r-mode frequencies for $\Omega \gtrsim 0.3 \Omega_K$. We also limited our minimum rotation rate at $\Omega = 0.7\Omega_K$ because of numerical reasons, as already stated in the introduction (lower rotation rates would demand longer simulations). In the next section we discuss the numerical limits on the magnetic field.

The final equations obtained via the linear combinations and redefinitions of variables can be found in the appendix \ref{A1}. In appendix \ref{A2} we present simplified equations obtained from the equations in appendix \ref{A1} by neglecting the coefficients of the order $\Omega H^{i}$ and $H^{2}$. These simplified equations can be useful for sufficiently weak magnetic fields and sufficiently slow rotation rates, where neglecting the $\Omega H^{i}$ and $H^{2}$ terms could be justified. We used also these equations to compute the r-mode frequencies and the results are compared in the figure \ref{fig:rmode1}.

\section{The results for the r-mode frequencies (for $\ell=m=2$)}

The r-modes ($\ell=m=2$) were computed using the equations \eqref{one1}-\eqref{seven1}. We solve the system of equations with a 2D Lax-Wendroff scheme with non-constant coefficients \cite{Mitchell}. We refer the reader to a previous work \cite{Chirenti} for further details on the numerical setup used for obtaining the r-mode frequency and eigenfunction. (In \cite{Chirenti} it was used for non-magnetized and differentially rotating stars.)

\begin{figure}[!htb]
\begin{center}
\includegraphics[angle=270,width=1\linewidth]{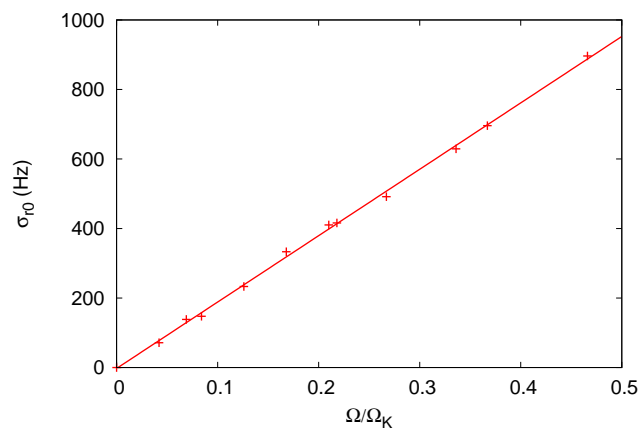}
\end{center}
\caption{The plot represents the r-mode frequency as a function of the rotation parameter for zero magnetic field.}
\label{fig:rmode0}
\end{figure}

We calculated the r-modes first for zero magnetic field. The dependence of the r-mode frequencies on the rotation parameter for the non-magnetized field case is shown in figure \ref{fig:rmode0}. We compared results with \cite{Ruoff} for the star with $\Omega/\Omega_K = 0.27$, and found that our results match with less than 3\% error. (For more results on r-modes of non-magnetized stars see also the papers \cite{Yoshida-Lee, Yoshida-Yoshida, Gaertig, Font, Kastaun}.)

\begin{figure}[!htb]
\begin{center}
\includegraphics[angle=270,width=1\linewidth]{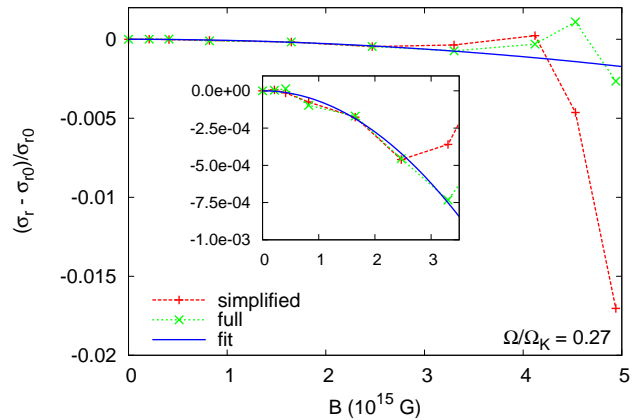}
\end{center}
\caption{The plot shows the relative change of the r-mode frequency as a function of the absolute value of the magnetic field. It compares the r-mode frequency computed using the equations  \eqref{one2}-\eqref{seven2}, (the equations with the linearized background coefficients), (the red line), with the r-mode obtained via the non-simplified equations \eqref{one1}-\eqref{seven1} (the green line). The solid line shows a quadratic fit done with the points before the breakup.}
\label{fig:rmode1}
\end{figure}

\begin{figure}[!htb]
\begin{center}
\includegraphics[angle=270,width=1\linewidth]{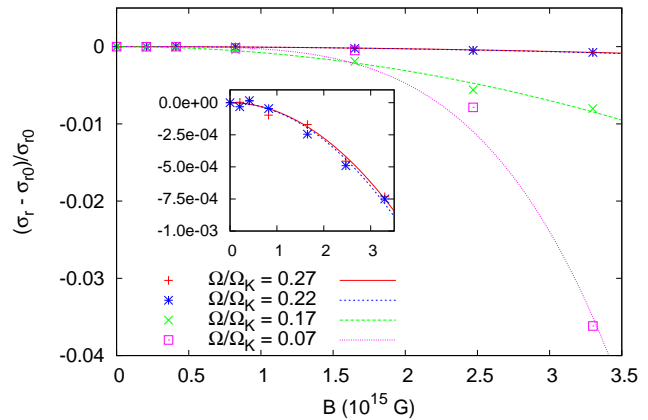}
\end{center}
\caption{The relative change of the r-mode frequency as a function of the absolute value of the magnetic field, for different values of the rotation parameter. (Computed with the full, non-simplified equations.) The lines show the quadratic (and quartic, for $\Omega/\Omega_K = 0.07$) fits.}
\label{fig:rmode2}
\end{figure}

In figures \ref{fig:rmode1} and \ref{fig:rmode2} we present r-modes as a function of magnetic field. For comparison we present in the figure \ref{fig:rmode1} also r-modes calculated via the simplified equations from the appendix \ref{A2} for the star with $\Omega/\Omega_K = 0.27$. The approximation of the simplified equations is shown to break down in this case at the value of magnetic field around $2.5\times 10^{15}$ G, while the results obtained from the full equations \emph{seem} to breakup at a larger magnetic field around $3.5\times 10^{15}$ G.  
For larger magnetic fields than $4\times 10^{15}$ G (where we do not entirely trust our results), we still see that the r-mode disappears completely due to the growth of another mode (possibly an Alfv\'en mode). We believe that the breakdown in the behavior of the r-mode frequencies is caused by the growth of this other mode and the subsequent deformation of the r-mode. This is consistent with the expectations based on the results of \cite{Rezzolla1, Rezzolla2, Rezzolla3, Lander}. 
 
As can be seen in the figure \ref{fig:rmode2}, the r-mode frequencies change very little when one turns on the magnetic field.  This is consistent with the observation of \cite{Lee} for Newtonian stars. The change of the frequencies is more pronounced for smaller values of the rotation parameter. For $\Omega/\Omega_{K}=0.07$ and magnetic field $3.3\times 10^{15}$ G, the r-mode frequency changes by a little less than 4\%. In case of larger rotation $\Omega/\Omega_{K}=0.17$ the same value of magnetic field changes the r-mode frequency by 1\%.  

Even though the variations are small, one can still clearly observe from Fig.\ref{fig:rmode2} the behavior of the frequencies with respect to the increasing magnetic field. Such a behavior seems to have remarkable features: the r-mode frequencies behave for sufficiently large $\Omega$ ($\Omega\sim 0.17 \Omega_{K}$) as $\sim B^{2}$, whereas for a smaller value of $\Omega$, ($\Omega/\Omega_{K}\approx 0.07$), the behavior of the frequencies with respect to the magnetic field is given as $\sim B^{4}$. Let us note that the $\sim B^{4}$ dependence  was observed for the r-modes of the spherical shell in \cite{Abbassi}. 

\begin{figure*}[!htb]
\begin{center}
\includegraphics[angle=270,width=0.7\linewidth]{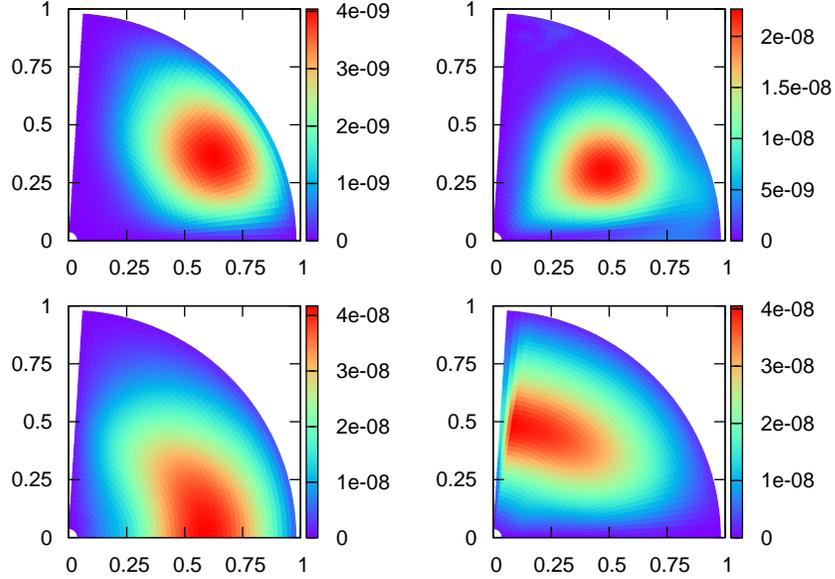}
\end{center}
\caption{The r-mode eigenfunctions for $3.3\times 10^{15}G$ magnetic field (at the center of the star) and the rotation rate $\Omega/\Omega_{K}=0.27$. The eigenfunctions go clockwise from the left upper corner as $\delta p, \delta u^{r}, \delta u^{\phi}, \delta u^{\theta}$. Also the vertical axis is the axis of rotation and the horizontal axis is lying in the equatorial plane.}
\label{fig:eigenf}
\end{figure*}

\begin{figure*}[!htb]
\begin{center}
\includegraphics[angle=270,width=0.315\linewidth]{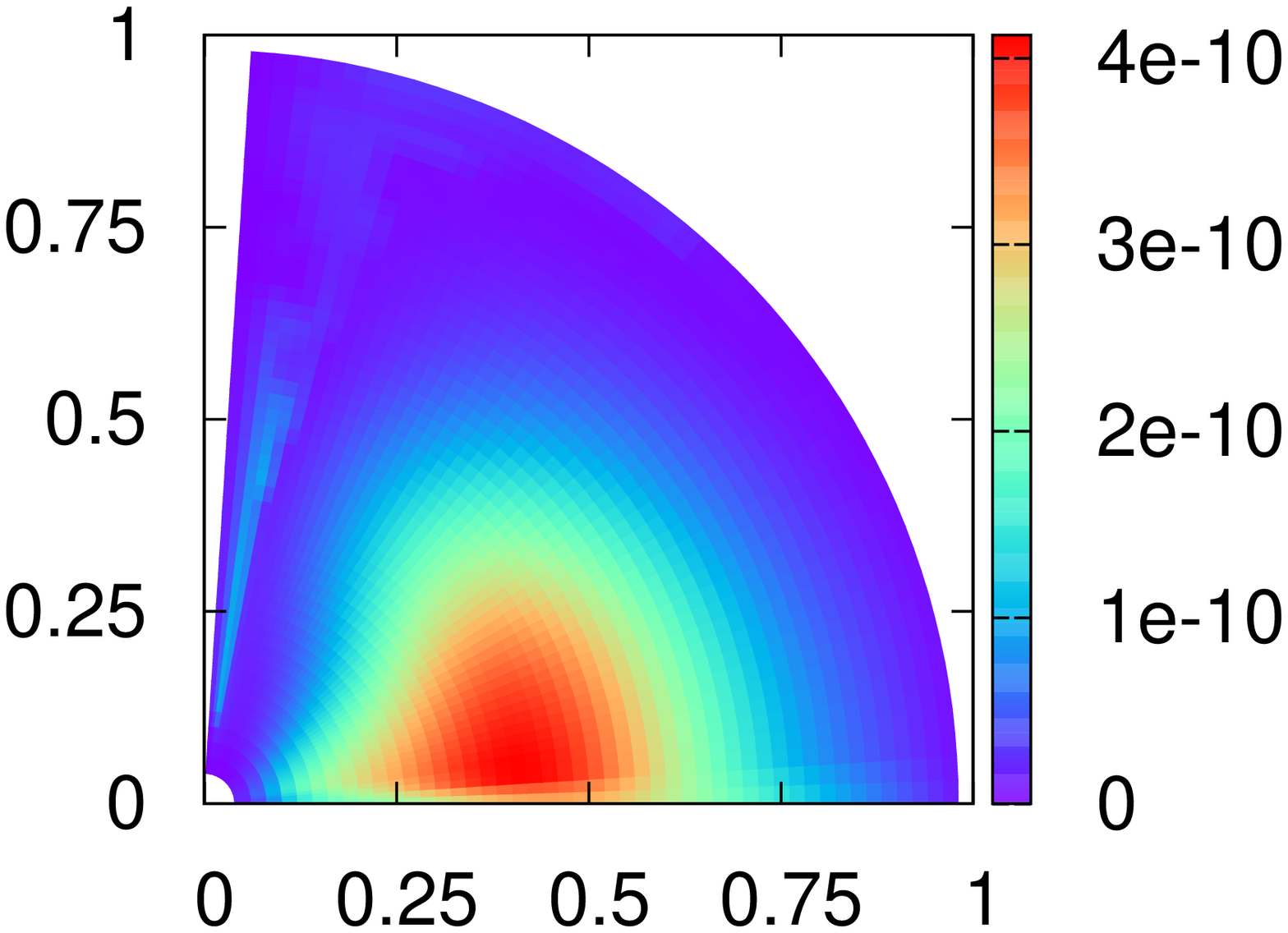}
\includegraphics[angle=270,width=0.315\linewidth]{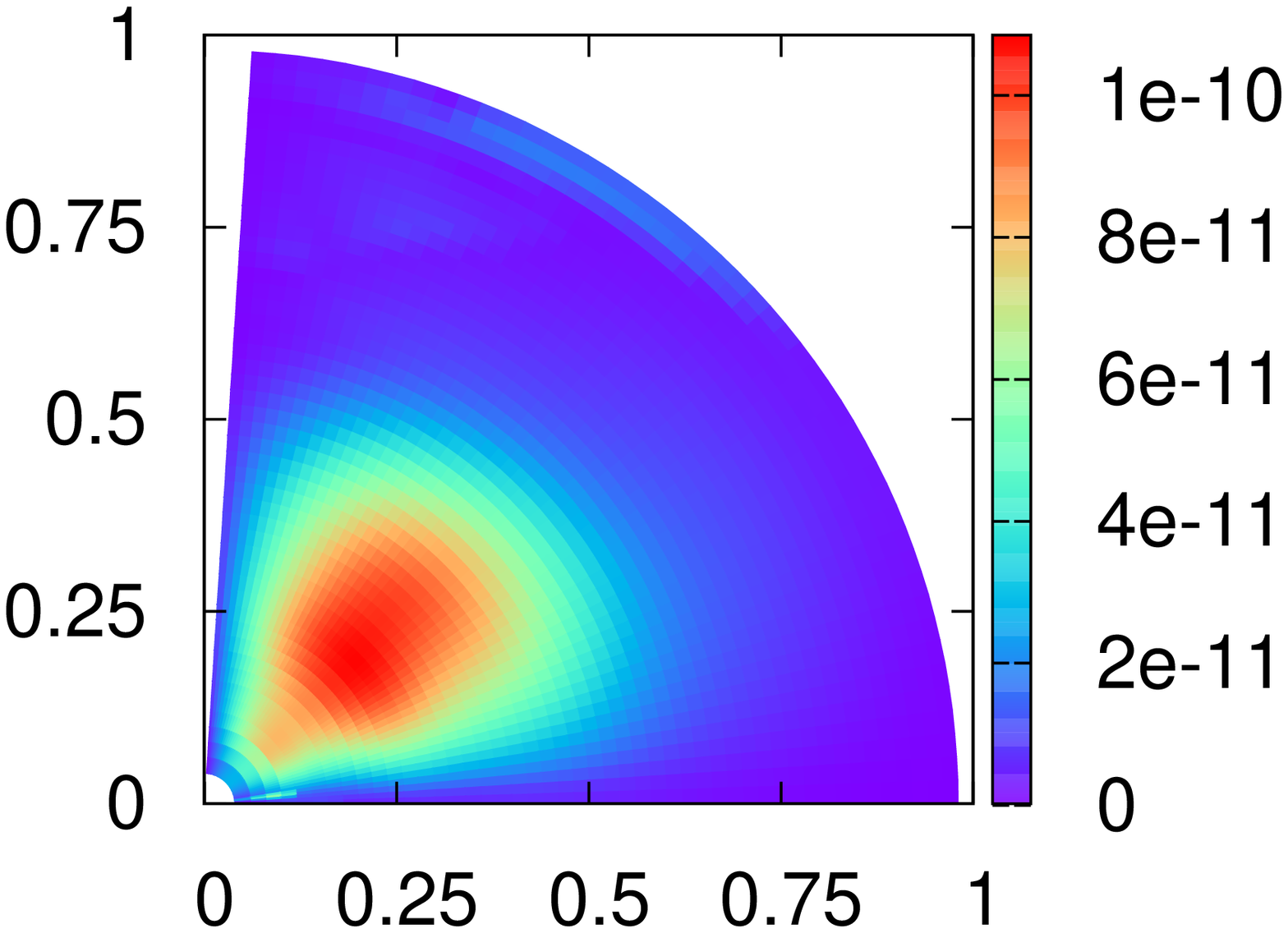}
\includegraphics[angle=270,width=0.315\linewidth]{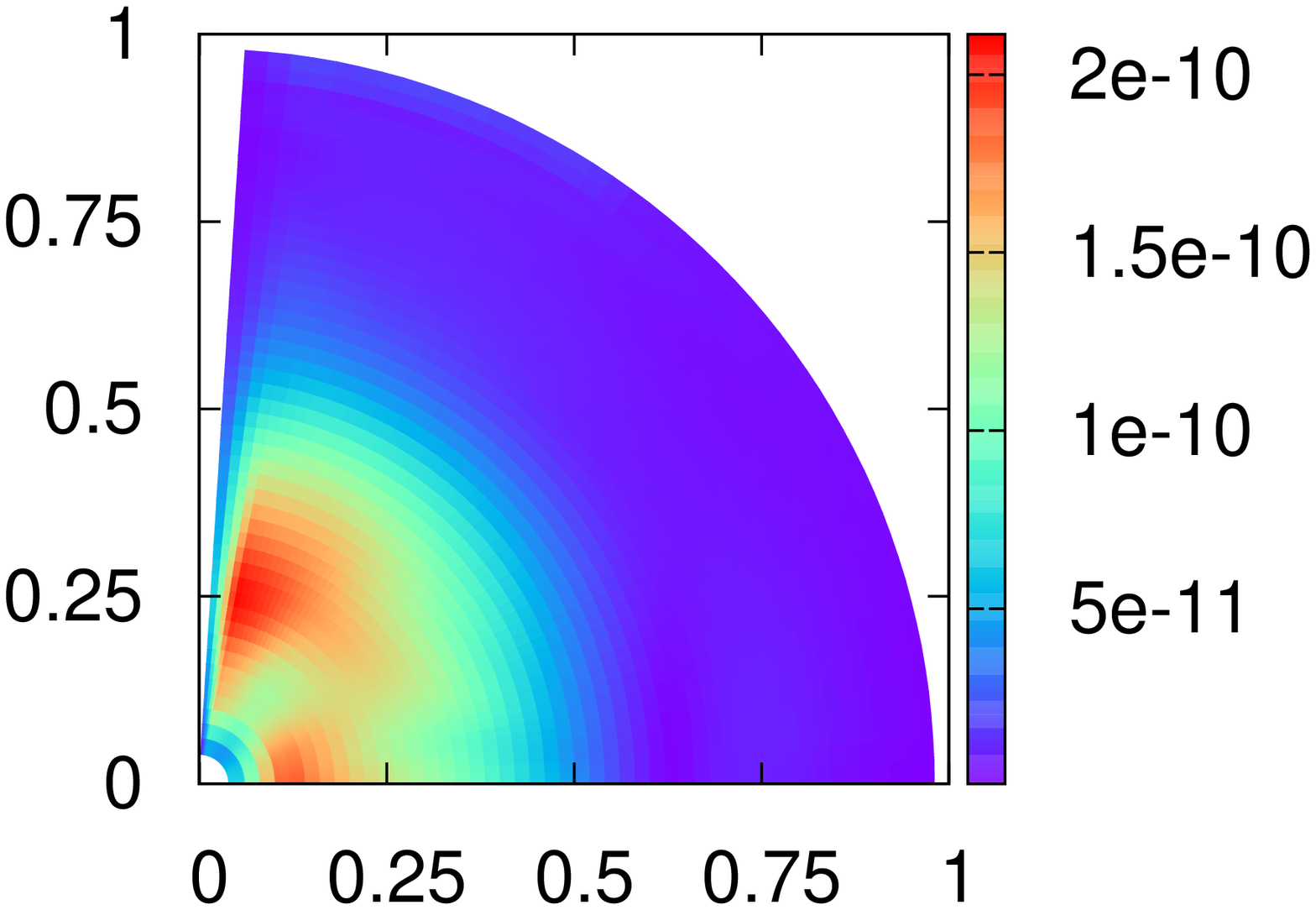}
\end{center}
\caption{The r-mode eigenfunctions for $3.3\times 10^{15}G$ magnetic field (at the center of the star) and the rotation rate $\Omega/\Omega_{K}=0.27$. The eiegenfunctions go from left to right as $\delta H^{r}, \delta H^{\theta}, \delta H^{\phi}$. Also the vertical axis is the axis of rotation and the horizontal axis is lying in the equatorial plane.}
\label{fig:eigenfB}
\end{figure*}

In figures \eqref{fig:eigenf} and \eqref{fig:eigenfB} we show the plots of the r-mode eigenfunctions for all the variables. (The eigenfunctions are shown in the plane given by the rotational axis and a perpendicular axis to the rotation.) The complicated interplay between r-modes and magnetic fields is more visible in the $\delta H^{\phi}$ eigenfunction, where we can see a sort of double peak. This happens for all rotation rates, and it is more pronounced for larger magnetic fields. We believe that this shows the deformation of the r-mode eigenfunction caused by other modes excited for large enough magnetic fields. For more details on the procedure used for extracting the eigenfunctions, see again \cite{Chirenti}.

\section{The r-mode instability and gravitational radiation}

The r-mode instability growth times (for $\ell=m=2$) are calculated by using the usual quadrupole formula (for the details see for example \cite{Andersson}). The characteristic timescale is calculated from the equation:
\begin{equation}
\frac{dE}{dt}=-\frac{2E}{t_{gw}}=-\frac{\int \rho|\delta v|^{2}dV}{t_{gw}}
\end{equation}
with $\delta v^{i}=\delta u^{i}/u^{t}$, ~$i=r,\theta,\phi$.
The energy time derivative is calculated from the quadrupole formula as:
\begin{equation*}
\frac{dE}{dt}|_{gw}=-(\sigma+m\Omega)\sum_{\ell=2}^{\infty}N_{\ell}\sigma^{2\ell+1}
(|\delta D_{\ell m}|^{2}+|\delta J_{\ell m}|^{2}),
\end{equation*}
with 
\[N_{\ell}=4\pi\frac{(\ell+1)(\ell+2)}{\ell(\ell-1)[(2\ell+1)!!]^{2}},\]
where $\delta D_{\ell m}$ and $\delta J_{\ell m}$ are the mass and the current multipoles defined as in \cite{Andersson}. (We use both mass and current multipoles, but because the r-modes involve only a perturbed velocity field, to the lowest order in $\Omega$, gravitational radiation through current multipoles dominates over that produced by mass multipoles \cite{Andersson, Lindblom2}.)
 In particular the multipoles can be expressed as:
\begin{equation*}
\delta D_{\ell m}=\int \delta \rho r^{\ell}Y^{*}_{\ell m}dV
\end{equation*}
and
\begin{equation*}
\delta J_{\ell m}=2\frac{\ell}{\ell+1}\int r^{\ell}(\rho\delta v^{i}+\delta \rho v^{i})Y^{B*}_{i~ \ell m}dV.
\end{equation*}
(Here $Y_{\ell m}$ and $Y^{B i}_{\ell m}$ are the multipoles defined in \cite{Thorne}.)

\begin{figure}[!htb]
\begin{center}
\includegraphics[angle=270,width=1\linewidth]{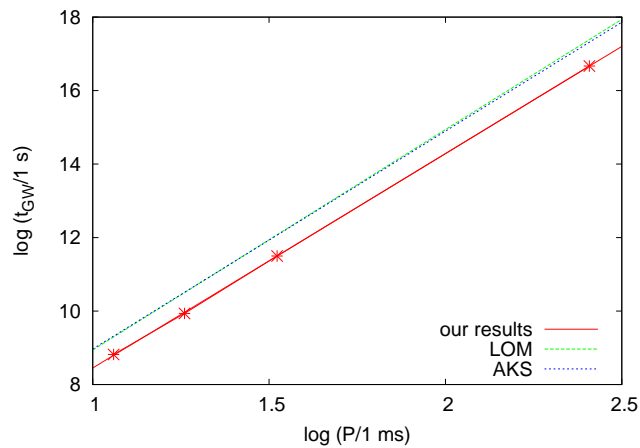}
\end{center}
\caption{The (normalized) logarithmic timescale for the r-mode instability growth as a function of the (normalized) logarithm of the period of rotation (for zero magnetic field). We compare here our results with the results of \cite{Andersson, Lindblom2}.}
\label{fig:logtau}
\end{figure}

We were able to fit the function $t_{gw}$ for zero magnetic fields as a functions of the rotation period $P$ as 
\[t_{gw}=\tau_{gw}(P/1\rm{ms})^{p_{gw}} \rm{s},\]
 with the dimensionless parameters $\tau_{gw}$ and $p_{gw}$ taking the values ~$\tau_{gw}=13.65$~ and ~$p_{gw}=5.83$. We compare our values for $\tau_{gw}$ and $p_{gw}$ with the values obtained in \cite{Lindblom, Andersson} in table \ref{tab:par} (see also figure \ref{fig:logtau}).

\begin{table}[h]
\caption{The $\tau_{gw}$ and $p_{gw}$ parameters for the case of zero magnetic fields. We are comparing our results with \cite{Andersson, Lindblom2} where the calculations were done for the Newtonian polytropes with stellar parameters close to ours. We obtained 27-34 \% faster emission of gravitational waves compared to the Newtonian setting.}
\label{tab:par}
\begin{ruledtabular}
\begin{tabular}{cccc}
 & our result & ref. \cite{Lindblom2} & ref. \cite{Andersson} \\	
\hline
$\tau_{gw}$  & 13.65 &   18.91 &       20.83 \\ 
\hline
$p_{gw}$  &  5.83 & 6 & 5.93\\  
\end{tabular}
\end{ruledtabular}
\end{table}

\begin{figure}[!htb]
\begin{center}
\includegraphics[angle=270,width=1\linewidth]{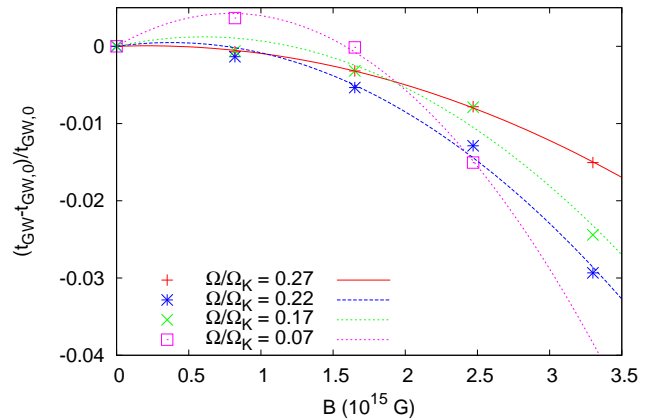}
\end{center}
\caption{The relative difference in the timescale for the r-mode instability growth as a function of the magnetic field.}
\label{fig:tauB1}
\end{figure}

The instability growth time scale $t_{gw}$ relative change due to the magnetic field is shown in figure \eqref{fig:tauB1}. We can see that the relative change becomes positive for lower values of magnetic fields (increasing the growth time and slowing down the emission of gravitational waves) and then becomes negative for larger values of the magnetic field (with the opposite effect), causing a relative change of up to $\sim$ 5\%. Similarly to the r-mode frequencies, the relative effect of the magnetic field is more pronounced for the lower rotation rates. However, to estimate the amount of gravitational waves emitted and the window of the instability we would need to calculate the viscosity damping rates and that is left for future work.

\section{Conclusions}
We presented here a model for a rotating magnetized star, in which we neglect the distortion of both the geometry and the fluid by a \emph{dipolar} magnetic field. We derived the full perturbation equations in the Cowling approximation. After solving the 2D time evolution problem with a Lax-Wendroff method, we computed the r-mode frequencies using the Fourier spectrum of the solution and we were able to extract the eigenfunctions of the perturbations. The frequencies and eigenfunctions of the r-mode were then used to calculate the growth time scale due to gravitational radiation, using the Newtonian quadrupole formalism. 

We found that the effect of the magnetic field on the frequencies is very small, (up to 5\% for the lowest rotation rates). For lower rotation rates the frequencies follow a $B^{4}$ dependence and, for higher rotation rates, a $B^{2}$ dependence. The effect on the r-mode growth time $t_{gw}$ indicates a faster emission of gravitational waves, compared to the Newtonian non-magnetized calculations of \cite{Lindblom2, Andersson}. We found that $t_{gw}$ is more significantly affected by the presence of general relativity ($\sim 30$\%) and less significantly by the presence of magnetic field (up to $\sim 5$\%). 

Our results indicate that the relative effects of the magnetic field are more pronounced for more slowly rotating stars. Therefore it could be possible that they achieve higher values for magnetars, that have rotation periods about 1000 times larger than the ones considered here (due to numerical limitations). However, it is not trivial to estimate how large these corrections would be, given the complicated dependence on both rotation and magnetic field of the solutions.

The effect of viscosity will play of course a key role in determining the actual instability window and is left for the future work. A more realistic description of the star would also need to include work with realistic equations of state and a stellar crust, together with considering the backreaction from the production of toroidal magnetic field \cite{Cuofano}. This is also left for the future.

\begin{acknowledgments}
The authors are especially thankful to Luciano Rezzolla and Shin'ichirou Yoshida for many useful discussions in different stages of this project. This work was supported by FAPESP and the Max Planck Society.
\end{acknowledgments}

\appendix

\section{The form of equations suitable for the numerical code}\label{A1}

The final dynamical equations for the numerical evolution are given as:

\begin{widetext}

\begin{eqnarray}\label{one1}
(p+\epsilon)  \left[  \Omega-(\Omega-\omega)\frac{\Gamma p}{\epsilon+p+H^{2}}  \right]  \delta p_{,\phi}-r^{2}\sin^{2}(\theta)(\Omega-\omega)\frac{\Gamma p}{\epsilon+p+H^{2}}   \left[   -H^{r}\delta \tilde H^{\phi}_{,r}-H^{\theta}\delta\tilde H^{\phi}_{,\theta}-\right.~~~~~~~~~~~~~~~~\nonumber\\
  \left.   \left(    H^{r}   \left[   \nu_{,r}+\frac{\nu_{,r}}{2}\frac{\epsilon+p}{\Gamma p}+\frac{2}{r}   \right]   +2\cot(\theta)H^{\theta}   \right)   \delta\tilde H^{\phi}+\frac{e^{\lambda}H^{r}}{r^{2}\sin^{2}(\theta)}\delta \tilde H^{r}_{,\phi}+\frac{H^{\theta}}{\sin^{2}(\theta)}\delta\tilde H^{\theta}_{,\phi}   \right]      +~~~~~~~~~~~~~~~~~~~~~~~~~~~~~~~~~~~~\\
+(p+\epsilon)\delta p_{,t}+e^{\nu/2}\Gamma p(\delta\tilde u^{r}_{,r}+\delta\tilde u^{\theta}_{,\theta}+\delta\tilde u^{\phi}_{,\phi})+e^{\nu/2}\Gamma p\left(\frac{\lambda_{,r}}{2}+\nu_{,r}+\frac{2}{r}\right)\delta \tilde u^{r}+~~~~~~~~~~~~\nonumber\\
+e^{\nu/2}\Gamma p\cot(\theta)\delta\tilde u^{\theta}=0~~~~~~~~~~~~~~~~~~~\nonumber
\end{eqnarray}

\begin{eqnarray}\label{two1}
\left[    e^{-\nu/2}\Omega-\frac{e^{\lambda}H^{r 2}(\Omega-\omega)}{\epsilon+p+H^{2}}   \right]    \delta\tilde H^{r}_{,\phi}-\frac{H^{r}r^{2}\sin^{2}(\theta)(\Omega-\omega)}{\epsilon+p+H^{2}}   \left[   \frac{p+\epsilon}{r^{2}\sin^{2}(\theta)}\delta p_{,\phi}-H^{r}\delta\tilde H^{\phi}_{,r}-H^{\theta}\delta\tilde H^{\phi}_{,\theta}- \right.~~~~~~~~~~~~~~~~~~~~~~~~~~~~~~\nonumber\\
  \left.  \left(    H^{r}   \left[   \frac{\nu_{,r}}{2}\frac{p+\epsilon}{\Gamma p}+\nu_{,r}+\frac{2}{r}   \right]   +2\cot(\theta)H^{\theta}   \right)  \delta\tilde H^{\phi}+\frac{H^{\theta}}{\sin^{2}(\theta)}\delta\tilde H^{\theta}_{,\phi}\right]  +~~~~~~~~~~~~~~~~~~~~~~~~~~~~~~~~~~~~\\
+e^{-\nu/2}\delta \tilde H^{r}_{,t}+[H^{r}_{,\theta}+H^{r}\cot(\theta)]\delta\tilde u^{\theta}+H^{r}(\delta\tilde u^{\phi}_{,\phi}+\delta\tilde u^{\theta}_{,\theta})-(H^{\theta}_{,\theta}+\cot(\theta)H^{\theta})\delta\tilde u^{r}-H^{\theta}\delta\tilde u^{r}_{,\theta}=0.~~~~~~~~~~~~~~~~~~~~~~~~~~~\nonumber 
\end{eqnarray}

\begin{eqnarray}\label{three1}
\left[\Omega e^{-\nu/2}-\frac{(\Omega-\omega)H^{\theta 2}r^{2}}{\epsilon+p+H^{2}}\right]\delta\tilde H^{\theta}_{,\phi}-\frac{(\Omega-\omega)r^{2}\sin^{2}(\theta)H^{\theta}}{\epsilon+p+H^{2}}\left[ \frac{p+\epsilon}{r^{2}\sin^{2}(\theta)}\delta p_{,\phi}-H^{r}\delta\tilde H^{\phi}_{,r}-H^{\theta}\delta\tilde H^{\phi}_{,\theta}-\right.~~~~~~~~~~~~~~~~~~~~~~~~~~~~~~~~~~~~~\nonumber\\
\left.\left(H^{r}\left[\frac{\nu_{,r}}{2}\frac{p+\epsilon}{\Gamma p}+\nu_{,r}+\frac{2}{r}\right]+2\cot(\theta)H^{\theta}\right)\delta\tilde H^{\phi}+\frac{e^{\lambda} H^{r}}{r^{2}\sin^{2}(\theta)}\delta\tilde H^{r}_{,\phi}\right]+~~~~~~~~~~~~~~~~~~~~~~~~~~~~~\\
+e^{-\nu/2}\delta\tilde H^{\theta}_{,t}+H^{\theta}(\delta\tilde u^{r}_{,r}+\delta\tilde u^{\phi}_{,\phi})+\left[H^{\theta}_{,r}+H^{\theta}\left(\frac{\nu_{,r}\frac{p+\epsilon}{\Gamma p}+\lambda_{,r}}{2}+\nu_{,r}+\frac{2}{r}\right)\right]\delta\tilde u^{r}-~~~~~~~~~~~~~~~~~~~~~~~~~~~~~~~~~~~~~~~~~~~~\nonumber\\
\left[H^{r}_{,r}+H^{r}\left(\frac{\nu_{,r}\frac{p+\epsilon}{\Gamma p}+\lambda_{,r}}{2}+\nu_{,r}+\frac{2}{r}\right)\right]\delta\tilde u^{\theta}-H^{r}\delta\tilde u^{\theta}_{,r}=0.~~~~~~~~~~~~~~~~~~~~~~~~~~~~~~~\nonumber
\end{eqnarray}

\begin{eqnarray}\label{four1}
\left[       \Omega(\epsilon+p+H^{2})-(\Omega-\omega)(H^{2}+\Gamma p)         \right]        \delta\tilde u^{\phi}_{,\phi}+~~~~~~~~~~~~~~~~~~~~~~~~\nonumber\\
+   \left[    -(\epsilon+p+r^{2}H^{\theta 2})\omega_{,r}+2\omega e^{\lambda}H^{r}      \left(       \frac{H^{r}}{r}+\cot(\theta)H^{\theta}      \right)    +    \right.~~~~~~~~~~~~~~~~~~~~~~~~~~~~\nonumber\\
          +\Omega e^{\lambda}          \left(         H^{r 2}        \left[        \lambda_{,r}+\frac{\nu_{,r}}{2}\frac{(\Gamma+1)p+\epsilon}{\Gamma p}        \right]      +H^{r}H^{r}_{,r}+H^{\theta}H^{r}_{,\theta}    \right) +~~~~~~~~~~~~~~~~~~~~~\nonumber\\
+(\Omega-\omega)   \left\{   (\epsilon+p+H^{2})   \left(   \frac{2}{r}-\nu_{,r}   \right)  -  \left(  \frac{\lambda_{,r}}{2}+\frac{2}{r}   \right)  (H^{2}+\Gamma p)-\nu_{,r}(r^{2}H^{\theta 2}+\Gamma p)-  \right.~~~~~~~~\nonumber\\
\frac{\nu_{,r}}{2}r^{2}H^{\theta 2}\frac{p+\epsilon}{\Gamma p}-  \left.   \left.\left(   e^{\lambda}H^{r}H^{r}_{,r}+r^{2}H^{\theta}H^{\theta}_{,\theta}   \right)  +\nu_{,r}H^{r 2}e^{\lambda}  \right\}  \right]  \delta\tilde u^{r}+~~~~~~~~~~~~~~~~~\nonumber\\
+\left[   (\Omega-\omega)\left(    \{2(\epsilon+p)+H^{2}-\Gamma p\}\cot(\theta)+r^{2}H^{r}H^{\theta}\frac{\nu_{,r}}{2}\frac{(3\Gamma +1)p+\epsilon}{\Gamma p}    \right)\right.+~~~~~~~~\\
\left.+\Omega r^{2}H^{r}H^{\theta}\frac{\nu_{,r}}{2}    \left(   1+\frac{\epsilon+p}{\Gamma p}   \right)   +\omega r^{2}    \left[   2H^{\theta}   \left(   \frac{2H^{r}}{r}+H^{\theta}\cot(\theta)    \right)    +H^{r}H^{\theta}_{,r}+H^{\theta}H^{\theta}_{,\theta}   \right]    +r^{2}H^{\theta}H^{r}\omega_{,r}    \right]    \delta\tilde u^{\theta}+~~~~~~~~~~~~~\nonumber\\
\left[(2\Omega-\omega)r^{2}H^{\theta 2}-(\Omega-\omega)\left(H^{2}+\Gamma p\right)\right]\delta\tilde u^{\theta}_{,\theta}+(2\Omega-\omega)e^{\lambda}H^{r}H^{\theta}\delta\tilde u^{r}_{,\theta}+~~~~~~~~~~~~~~\nonumber\\
+(2\Omega-\omega)r^{2}H^{r}H^{\theta}\delta\tilde u^{\theta}_{,r}+~~~~~~~~~~~~~~~~~~~\nonumber\\
+\left[     (2\Omega-\omega) e^{\lambda}H^{r 2}-(\Omega-\omega)     \left(     \Gamma p+H^{2}     \right)     \right]   \delta\tilde u^{r}_{,r}+
(\epsilon+p+H^{2})\delta\tilde u^{\phi}_{,t}+e^{\nu/2}   \left[    \frac{p+\epsilon}{r^{2}\sin^{2}(\theta)}\delta p_{,\phi}-\right.~~~~~~~~~~~~\nonumber\\
H^{r}\delta\tilde H^{\phi}_{,r}-H^{\theta}\delta\tilde H^{\phi}_{,\theta}-  \left.    \left[      \left(         \frac{\nu_{,r}}{2}+\frac{2}{r}           \right)    H^{r}+2\cot(\theta)H^{\theta}        \right]         \delta\tilde H^{\phi}+\frac{e^{\lambda}H^{r}}{r^{2}\sin^{2}(\theta)}\delta\tilde H^{r}_{,\phi}+\frac{H^{\theta}}{\sin^{2}(\theta)}\delta\tilde H^{\theta}_{,\phi}    \right]   =0.~~~~~~~~~~~\nonumber
\end{eqnarray}

\begin{eqnarray}\label{five1}
\Omega (\epsilon+p+H^{2})e^{-\nu/2}\delta u^{r}_{,\phi}+\frac{(\epsilon+p+H^{2})e^{-\nu/2}\Omega r^{2}H^{r}H^{\theta}}{\epsilon+p+e^{\lambda}H^{r 2}}\delta u^{\theta}_{,\phi}+~~~~~~~~~~~~~~\nonumber\\
+e^{-\nu/2}r^{2}\sin^{2}(\theta)(\epsilon+p+H^{2})    \left[    e^{-\lambda}     \left(    \omega_{,r}+    \left\{    \nu_{,r}-\frac{2}{r}   \right\}   (\Omega-\omega)    \right) \right. -~~~~~~~~~~\nonumber\\
\frac{\Omega-\omega}{\epsilon+p+e^{\lambda}H^{r 2}}   \left.   \left\{   \frac{\nu_{,r}H^{r 2}}{2}\frac{(2\Gamma+1)p+\epsilon}{\Gamma p}+2H^{r}H^{\theta}\cot(\theta)   \right\}   \right]  \delta\tilde u^{\phi}-~~~~~~~~\nonumber\\
\frac{H^{r}r^{2}\sin^{2}(\theta)e^{-\nu/2}(\Omega-\omega)(\epsilon+p+H^{2})}{\epsilon+p+e^{\lambda}H^{r 2}}    \left[     H^{r}\delta\tilde u^{\phi}_{,r}+H^{\theta}\delta\tilde u^{\phi}_{,\theta}    \right]   +~~~~~~~~~\nonumber\\
+\frac{e^{-\nu/2}(\epsilon+p)(\epsilon+p+H^{2})}{\epsilon+p+e^{\lambda}H^{r 2}}\delta\tilde u^{r}_{,t}+
\frac{e^{-\lambda}(p+\epsilon)\nu_{,r}}{2}\left[1+\frac{\epsilon+p}{\Gamma p}\right]\delta p+~~~~~~~~~~\\
+e^{-\lambda}(p+\epsilon)\delta p_{,r}-\frac{H^{r}(\epsilon+p+H^{2})}{\epsilon+p+e^{\lambda}H^{r 2}}(\delta\tilde H^{r}_{,r}+\delta\tilde H^{\phi}_{,\phi}+\delta\tilde H^{\theta}_{,\theta})-\frac{H^{\theta}(\epsilon+p)}{\epsilon+p+e^{\lambda}H^{r 2}}\delta\tilde H^{r}_{,\theta}+~~~~~~~~\nonumber\\
\left[e^{-\lambda}r^{2}H^{\theta}\left(\frac{\nu_{,r}}{2}+\frac{2}{r}\right)-\cot(\theta)H^{r}-\frac{r^{2}H^{\theta}\left\{H^{r}\left(H^{\theta}\cot(\theta)+H^{r}\left[
\frac{\nu_{r}}{2}+\frac{2}{r}\right]\right)+(\epsilon+p)e^{-\lambda}\frac{\nu_{,r}}{2}\frac{(\Gamma+1)p+\epsilon}{\Gamma p}\right\}}{\epsilon+p+e^{\lambda}H^{r 2}}\right] \delta\tilde H^{\theta}+~~~~~~~~\nonumber\\
+\frac{H^{\theta}e^{-\lambda}r^{2}(\epsilon+p)}{\epsilon+p+e^{\lambda}H^{r 2}}\delta\tilde H^{\theta}_{,r}-\frac{H^{r}(\epsilon+p+H^{2})}{\epsilon+p+e^{\lambda}H^{r 2}}\left(\frac{\nu_{,r}+\lambda_{,r}}{2}+\frac{\nu_{,r}}{2}\frac{p+\epsilon}{\Gamma p}+\frac{2}{r}\right)\delta\tilde H^{r}+\frac{H^{r}H^{\theta}(p+\epsilon)}{\epsilon+p+e^{\lambda}H^{r 2}}\delta p_{,\theta}=0~~~~~~~~\nonumber
\end{eqnarray}

\begin{eqnarray}\label{six1}
(\epsilon+p+H^{2})e^{-\nu/2}\Omega\delta\tilde u^{\theta}_{,\phi}+\frac{(\epsilon+p+H^{2})e^{-\nu/2}\Omega e^{\lambda}H^{\theta}H^{r}}{\epsilon+p+r^{2}H^{\theta 2}} \delta\tilde u^{r}_{,\phi}-~~~~~~~~~~~~~~~~~~~~~~~~~~~~~~~~~~~~~~~~~~~\nonumber\\
(\epsilon+p+H^{2})\sin^{2}(\theta)e^{-\nu/2}    \left[    2\cot(\theta)(\Omega-\omega)+\frac{H^{r}H^{\theta}r^{2}(\Omega-\omega)\nu_{,r}}{2(\epsilon+p+r^{2}H^{\theta 2})}\frac{(2\Gamma+1)p+\epsilon}{\Gamma p}-     \right.~~~~~~~~~~~~~~~~~~~~~~~~~~~~~~~~~~~~~~~~~~\nonumber\\
\frac{H^{\theta}H^{r}r^{2}}{\epsilon+p+r^{2}H^{\theta 2}}   \left.   \left(   \omega_{,r}+    \left\{   \nu_{,r}-\frac{2}{r}   \right\}(\Omega-\omega)   \right)    \right]    \delta\tilde u^{\phi}-~~~~~~~~~~~~~~~~~~~~~~~~~~~~~~~~~~\nonumber\\
\frac{H^{\theta}e^{-\nu/2}r^{2}\sin^{2}(\theta)(\Omega-\omega)(\epsilon+p+H^{2})}{\epsilon+p+r^{2}H^{\theta 2}}      \left[     H^{r}\delta\tilde u^{\phi}_{,r}+H^{\theta}\delta\tilde u^{\phi}_{,\theta}   \right]+~~~~~~~~~~~~~~~~~~~~~~~~~~~~~~~~~~~~~~~~~~~~\nonumber\\
+\frac{e^{-\nu/2}(\epsilon+p)(\epsilon+p+H^{2})}{\epsilon+p+r^{2}H^{\theta 2}}\delta\tilde u^{\theta}_{,t}+\frac{\nu_{,r}H^{r}H^{\theta}(p+\epsilon)}{2(\epsilon+p+r^{2}H^{\theta 2})}\left[1+\frac{\epsilon+p}{\Gamma p}\right]\delta p+~~~~~~~~~~~~~~~~~~~~~~~~~~~~~~~~~~~\\
+\frac{H^{\theta}H^{r}(p+\epsilon)}{\epsilon+p+r^{2}H^{\theta 2}}\delta p_{,r}-\frac{H^{\theta}(\epsilon+p+H^{2})}{\epsilon+p+r^{2}H^{\theta 2}}(\delta\tilde H^{r}_{,r}+\delta\tilde H^{\phi}_{,\phi}+\delta\tilde H^{\theta}_{,\theta})+\frac{e^{\lambda}H^{r}(\epsilon+p)}{r^{2}(\epsilon+p+r^{2}H^{\theta 2})}\delta\tilde H^{r}_{,\theta}+~~~~~~~~~~~~~~~~~~~~\nonumber\\
+\left[         \frac{H^{r}}{\epsilon+p+r^{2}H^{\theta 2}}         \left(         r^{2}H^{\theta 2}       \left\{        \frac{\nu_{,r}}{2}+\frac{2}{r}     \right\}        -e^{\lambda}\cot(\theta)H^{r}H^{\theta}-(\epsilon+p)\frac{\nu_{,r}}{2}\frac{(\Gamma+1)p+\epsilon}{\Gamma p}      \right) \right. -~~~~~~~~~~~~~~~~~~~~~~~~~~~~~~~~~~~\nonumber\\
H^{\theta}\cot(\theta)-H^{r} \left.   \left\{\frac{\nu_{,r}}{2}+\frac{2}{r}\right\}\right]\delta\tilde H^{\theta}-~~~~~~~~~~~~~~~~~~~~~~~~~~~~~~~~~~~~~~~~~\nonumber\\
\frac{H^{r}(\epsilon+p)}{\epsilon+p+r^{2}H^{\theta 2}}\delta\tilde H^{\theta}_{,r}-\frac{H^{\theta}(\epsilon+p+H^{2})}{\epsilon+p+r^{2}H^{\theta 2}}\left[\frac{\nu_{,r}+\lambda_{,r}}{2}+\frac{\nu_{,r}}{2}\frac{p+\epsilon}{\Gamma p}+\frac{2}{r}\right]\delta\tilde H^{r}+\frac{p+\epsilon}{r^{2}}\delta p_{,\theta}=0~~~~~~~~~~~~~~~~~~~~~~~~~~~~~~~~~~\nonumber
\end{eqnarray}

\begin{eqnarray}\label{seven1}
\Omega\left[\epsilon+p-H^{2}\right]\delta\tilde H^{\phi}_{,\phi}+\Omega H^{r}\frac{\nu_{,r}(p+\epsilon)}{2}\left(1+\frac{\epsilon+p}{\Gamma p}\right)\delta p+\Omega H^{r}(p+\epsilon)\delta p_{,r}-~~~~~~~~~\nonumber\\
\Omega H^{2}\left(\delta\tilde H^{r}_{,r}+\delta\tilde H^{\theta}_{,\theta}\right)-\Omega \left(\frac{\nu_{,r}+\lambda_{,r}}{2}+\frac{\nu_{,r}}{2}\frac{p+\epsilon}{\Gamma p}+\frac{2}{r}\right)H^{2}\delta\tilde H^{r}-\Omega \cot(\theta)H^{2}\delta\tilde H^{\theta}-~~~~~~~~~~\\
\Omega H^{\theta}(p+\epsilon)\delta p_{,\theta}+(\epsilon+p)e^{\nu/2}\left[ e^{-\nu/2}\delta\tilde H^{\phi}_{,t}-\left[\frac{1}{2}\frac{p+\epsilon}{\Gamma p}+1\right]\nu_{,r}H^{r}\delta\tilde u^{\phi}-H^{r}\delta\tilde u^{\phi}_{,r}-H^{\theta}\delta\tilde u^{\phi}_{,\theta}\right]=0.~~~~~~~~~~~\nonumber
\end{eqnarray}

\section{The equations with the linearized coefficients}\label{A2}

\begin{eqnarray}\label{one2}
\left[   \Omega(\epsilon+p)-(\Omega-\omega)\Gamma p   \right]  \delta p_{,\phi}  +~~~~~~~~~~~~~~~~~~~~~~~~~~~~~~~~~~~~~~~~~~~~~~~~~~~~~~~~~~~~~~~~\nonumber\\
+ (\epsilon+p)\delta p_{,t}+e^{\nu/2}\Gamma p(\delta\tilde u^{r}_{,r}+\delta\tilde u^{\theta}_{,\theta}+\delta\tilde u^{\phi}_{,\phi})+e^{\nu/2}\Gamma p   \left(    \frac{\lambda_{,r}}{2}+\nu_{,r}+\frac{2}{r}   \right)    \delta \tilde u^{r}+   ~~~~~~~~~~~~\\
+   e^{\nu/2}\Gamma p\cot(\theta)\delta\tilde u^{\theta}    =0~~~~~~~~~~~~~~~~~~~~~~~~~~~~~~~~~~~~~~~~~~~~~~~~~~\nonumber
\end{eqnarray}

\begin{eqnarray}\label{two2}
e^{-\nu/2}\Omega\delta\tilde H^{r}_{,\phi}+e^{-\nu/2}\delta \tilde H^{r}_{,t}+~~~~~~~~~~~~~~~~~~~~~~~~~~~~~~~~~~~~~~~~~~~~\nonumber\\
+[H^{r}_{,\theta}+H^{r}\cot(\theta)]\delta\tilde u^{\theta}+H^{r}(\delta\tilde u^{\phi}_{,\phi}+\delta\tilde u^{\theta}_{,\theta})-(H^{\theta}_{,\theta}+\cot(\theta)H^{\theta})\delta\tilde u^{r}-H^{\theta}\delta\tilde u^{r}_{,\theta}=0.~~~~~~~~~~~~~~~~~~~~~~~~~~~ 
\end{eqnarray}

\begin{eqnarray}\label{three2}
\Omega e^{-\nu/2}\delta\tilde H^{\theta}_{,\phi}+~~~~~~~~~~~~~~~~~~~~~~~~~~~~~~~~~~~~~~~~~~~~~~~~~~~~~~~~~~~~~~~~~~~~~~~~~~~~~~~~~~~~~~~~~~~~~~~~~~~~~~~~~~~~~~~~\nonumber\\
+e^{-\nu/2}\delta\tilde H^{\theta}_{,t}+H^{\theta}(\delta\tilde u^{r}_{,r}+\delta\tilde u^{\phi}_{,\phi})+\left[H^{\theta}_{,r}+H^{\theta}\left(\nu_{,r}\left\{\frac{p+\epsilon}{2\Gamma p}+1\right\}+\frac{\lambda_{,r}}{2}+\frac{2}{r}\right)\right]\delta\tilde u^{r}-~~~~~~~~~~~~~~~~~~~~~~~~~~~~~~~~~~~~~~~~~~~~\\
\left[H^{r}_{,r}+H^{r}\left(\nu_{,r}\left\{\frac{p+\epsilon}{2\Gamma p}+1\right\}+\frac{\lambda_{,r}}{2}+\frac{2}{r}\right)\right]\delta\tilde u^{\theta}-H^{r}\delta\tilde u^{\theta}_{,r}=0.~~~~~~~~~~~~~~~~~~~~~~~~~~~~~~~\nonumber 
\end{eqnarray}

\begin{eqnarray}\label{four2}
  \left[   \Omega(\epsilon+p)-(\Omega-\omega)\Gamma p)   \right]   \delta\tilde u^{\phi}_{,\phi}
+   ~~~~~~~~~~~~~~~~~~~~~\nonumber\\
+ \left[    -(\epsilon+p)\omega_{,r}+(\Omega-\omega)    \left\{    (\epsilon+p)    \left(    \frac{2}{r}-\nu_{,r}    \right)   -       \left(     \frac{\lambda_{,r}}{2}+\nu_{,r}+\frac{2}{r}    \right)    \Gamma p  \right\}  \right]  \delta\tilde u^{r}+~~~~~~~~~~~~\\
+      (\Omega-\omega)      \{2(\epsilon+p)-\Gamma p\}\cot(\theta)     \delta\tilde u^{\theta}-
    (\Omega-\omega)  \Gamma p     \delta\tilde u^{\theta}_{,\theta}-
(\Omega-\omega) \Gamma p     \delta\tilde u^{r}_{,r}+
(\epsilon+p)\delta\tilde u^{\phi}_{,t}+e^{\nu/2}     \left[     \frac{\epsilon+p}{r^{2}\sin^{2}(\theta)}\delta p_{,\phi}-     \right.     ~\nonumber\\
H^{r}\delta\tilde H^{\phi}_{,r}-H^{\theta}\delta\tilde H^{\phi}_{,\theta}-     \left.    \left[     \left(     \frac{\nu_{,r}}{2}+\frac{2}{r}     \right)     H^{r}+2\cot(\theta)H^{\theta}     \right]      \delta\tilde H^{\phi}+\frac{e^{\lambda}H^{r}}{r^{2}\sin^{2}(\theta)}\delta\tilde H^{r}_{,\phi}+\frac{H^{\theta}}{\sin^{2}(\theta)}\delta\tilde H^{\theta}_{,\phi}     \right]     =0.~~~~~~\nonumber
\end{eqnarray}

\begin{eqnarray}\label{five2}
\Omega (\epsilon+p)e^{-\nu/2}\delta u^{r}_{,\phi}+~~~~~~~~~~~~~~~~~~~~~~~~~~~~~~~~~~~~~~~~~~~~~~~~~~~\nonumber\\
+e^{-\nu/2}r^{2}\sin^{2}(\theta)  e^{-\lambda}(\epsilon+p)    \left(    \omega_{,r}+     \left\{     \nu_{,r}-\frac{2}{r}     \right\}     (\Omega-\omega)     \right)       \delta\tilde u^{\phi}+~~~~~~~~~~~~~~~~~~~~~~~~~~~~~~~~~~\nonumber\\
+e^{-\nu/2}(\epsilon+p)\delta\tilde u^{r}_{,t}+
(\epsilon+p)\frac{e^{-\lambda}\nu_{,r}}{2}     \left[  \frac{\epsilon+p}{\Gamma p} +1    \right]     \delta p+~~~~~~~~~~~~~~~~~~~~~~~~~~~~~~~~~~~\nonumber\\
+(\epsilon+p)e^{-\lambda}\delta p_{,r}-H^{r}(\delta\tilde H^{r}_{,r}+\delta\tilde H^{\phi}_{,\phi}+\delta\tilde H^{\theta}_{,\theta})-H^{\theta}\delta\tilde H^{r}_{,\theta}+~~~~~~~~~~~~~~~~~~~~~~~~~~~~~~~\\
 +   \left[        e^{-\lambda}r^{2}H^{\theta}     \left(      \frac{\nu_{,r}}{2}+\frac{2}{r}      \right)     -\cot(\theta)H^{r}-r^{2}H^{\theta}e^{-\lambda}\frac{\nu_{,r}}{2}\left\{\frac{\epsilon+p}{\Gamma p}+1\right\}      \right]     \delta\tilde H^{\theta}+~~~~~~~~~~~~~~~~~~~~~~~~~\nonumber\\
+H^{\theta}e^{-\lambda}r^{2}\delta\tilde H^{\theta}_{,r}-H^{r}     \left(      \frac{\lambda_{,r}}{2}+\frac{\nu_{,r}}{2}\left\{\frac{\epsilon+p}{\Gamma p}+1\right\}+\frac{2}{r}      \right)     \delta\tilde H^{r}=0~~~~~~~~~~~~~~~~~~~~~~~~~~~~~~~~~~\nonumber
\end{eqnarray}

\begin{eqnarray}\label{six2}
(\epsilon+p)e^{-\nu/2}\Omega\delta\tilde u^{\theta}_{,\phi}-~~~~~~~~~~~~~~~~~~~~~~~~~~~~~~~~~~~~~~~~~~~\nonumber\\
\sin^{2}(\theta)e^{-\nu/2}  (\epsilon+p)2\cot(\theta)(\Omega-\omega)   \delta\tilde u^{\phi}
+e^{-\nu/2}(\epsilon+p)\delta\tilde u^{\theta}_{,t}-~~~~~~~~~~~~~~~~~~~~~~~~~~~~~~~~~~~\nonumber\\
H^{\theta}(\delta\tilde H^{r}_{,r}+\delta\tilde H^{\phi}_{,\phi}+\delta\tilde H^{\theta}_{,\theta})+\frac{e^{\lambda}H^{r}}{r^{2}}\delta\tilde H^{r}_{,\theta}-~~~~~~~~~~~~~~~~~~~~~~~~~~~~~~~~~~~~~~~~\\
   \left[      H^{r}\frac{\nu_{,r}}{2}\left\{\frac{p+\epsilon}{\Gamma p}+1\right\}+H^{\theta}\cot(\theta)+H^{r}     \left\{     \frac{\nu_{,r}}{2}+\frac{2}{r}     \right\}     \right]     \delta\tilde H^{\theta}-~~~~~~~~~~~~~~~~~~~~~~~~~~~~~~~~~~~~~~~~~\nonumber\\
H^{r}\delta\tilde H^{\theta}_{,r}-H^{\theta}     \left[     \frac{\lambda_{,r}}{2}+\frac{\nu_{,r}}{2}\left\{\frac{p+\epsilon}{\Gamma p}+1\right\}+\frac{2}{r}     \right]     \delta\tilde H^{r}+\frac{\epsilon+p}{r^{2}}\delta p_{,\theta}=0~~~~~~~~~~~~~~~~~~~~~~~~~~~~~~~~~~\nonumber
\end{eqnarray}

\begin{eqnarray}\label{seven2}
\Omega\delta\tilde H^{\phi}_{,\phi}+\delta\tilde H^{\phi}_{,t}-e^{\nu/2}   \left(    \left[     \frac{1}{2}\frac{p+\epsilon}{\Gamma p}+1    \right]    \nu_{,r}H^{r}\delta\tilde u^{\phi}+H^{r}\delta\tilde u^{\phi}_{,r}+H^{\theta}\delta\tilde u^{\phi}_{,\theta}\right)=0.~~~~~~~~~~~~~~~~~~~~
\end{eqnarray}

\end{widetext}

\end{document}